\documentclass[twocolumn,reprint,aps]{revtex4}%
\usepackage{graphicx}
\usepackage{dcolumn}
\usepackage{bm}
\usepackage{color}
\usepackage{dcolumn}
\usepackage{amsmath}
\usepackage{multirow}
\usepackage{amsfonts}
\usepackage{amssymb}
\usepackage[colorinlistoftodos]{todonotes}%
\providecommand{\U}[1]{\protect\rule{.1in}{.1in}}
\fontfamily{cmss}
\providecommand{\U}[1]{\protect\rule{.1in}{.1in}}

\begin{document}
\title{Chiral charge-density wave in TiSe$_{2}$ due to photo-induced structural distortions}
\author{Darshana Wickramaratne}
\affiliation{Center for Computational Materials Science, US Naval Research Laboratory,
Washington, D.C. 20375, USA}
\author{R. D. Schaller}
\affiliation{Center for Nanoscale Materials, Argonne National Laboratory, Lemont, Illinois
60439, USA}
\author{G. P. Wiederrecht}
\affiliation{Center for Nanoscale Materials, Argonne National Laboratory, Lemont, Illinois
60439, USA}
\author{G. Karapetrov}
\affiliation{Department of Physics, Drexel University, Philadelphia, Pennsylvania 19104, USA}
\author{I. I. Mazin}
\affiliation{Department of Physics and Astronomy, George Mason University, Fairfax, VA
22030, USA}
\affiliation{Quantum Science and Engineering Center, George Mason University, Fairfax, VA
22030, USA}
\date{\today}

\begin{abstract}
A variety of experiments have been carried out to establish the origin of the
chiral charge-density wave transition in 1T-TiSe$_{2}$, which in turn
has led to contradictory conclusions on the origin of this transition. Some
studies suggest the transition is a phonon-driven structural distortion while
other studies suggest it is an excitonic insulator phase transition that is
accompanied by a lattice distortion. First, we propose these interpretations
can be reconciled if one analyzes the available experimental and theoretical
data within a formal definition of what constitutes an excitonic insulator as
initially proposed by Keldysh and Kopaev. Next, we present pump-probe
measurements of circularly polarized optical transitions and first-principles
calculations where we highlight the importance of accounting for structural
distortions to explain the finite chirality of optical transitions in the CDW
phase. We show that at the elevated electronic temperature that
occurs upon photoexcitation, there is a non-centrosymmetric structure that is
near-degenerate in energy with the centrosymmetric charge density wave
structure, which explains the finite chirality of the optical transitions
observed in the CDW phase of TiSe$_{2}$.

\end{abstract}
\maketitle

\section{Introduction}

TiSe$_{2}$ is claimed to exhibit signatures of two nontrivial phenomena: an
excitonic insulator (EI) phase, which is a Bose condensation of excitons, and
a chiral charge-density wave (CDW) phase, a state where time reversal symmetry
is spontaneously broken
\cite{kogar2017signatures,rohwer2011collapse,xu2020spontaneous,van2011chirality,hellgren2017critical,
weber2011electron,kidd2002electron,
rossnagel2002charge,cercellier2007evidence,ishioka2010chiral}. This has made
it the subject of extensive experimental and theoretical studies, which in
turn has led to diverging opinions on the microscopic nature of this
transition. Some authors have suggested that the CDW phase of TiSe$_{2}$ is
mostly due to an excitonic insulator transition and the structural distortion
follows this purely electronic transition \cite{monney2011exciton,
kogar2017signatures}, while other studies have argued that the CDW in
TiSe$_{2}$ is driven by electron-phonon
coupling~\cite{hughes1977structural,whangbo1992analogies,rossnagel2002charge}.
A third point of view has suggested that the combination of exciton
\textit{and} phonon interactions are required to explain the CDW phase
transition \cite{van2010exciton,zenker2013chiral}.

Indeed, some of the experimental and theoretical observations that led to
these conclusions don't seem mutually compatible. The plasmon softening
\cite{kogar2017signatures} and the insulating state observed in photoemission
measurements \cite{cercellier2007evidence} that were associated with the
condensation of excitons in the CDW phase can be reproduced with density
functional theory (DFT) calculations without invoking the role of excitons
\cite{lian2019charge, hellgren2017critical}. The chirality of the CDW phase
\cite{ishioka2010chiral} is also difficult to reconcile with the fact that the
CDW structure is centrosymmetric \cite{xu2020spontaneous}.

These contradictory viewpoints are understandable, given that there is no
distinct symmetry breaking associated with an EI transition that distinguishes
it from a structural phase transition. The distinction is solely in the eyes
of the beholder. An often discussed litmus test for an EI transition is a
\textit{gedanken experiment} where the nuclei are clamped to their equilibrium
positions, and the electron subsystem experiences a transition with the atomic
coordinates fixed in place. This is an unphysical criterion that neglects the
interaction between ions and electrons, which is present in any material.
Furthermore, this simplified consideration has a major conceptual problem: it
classifies any transition associated with a divergence of the one-electron
dielectric response, such as the well-known Peierls transition, as an EI, even
though such transitions are not usually described as condensation of excitons.

In this paper we will show that it is possible to reconcile these
contradictory viewpoints. Our primary focus will be to provide an explanation
for the finite chirality observed in the CDW phase of TiSe$_{2}$ and whether
this is associated with the condensation of excitons or not. We will do so by
presenting a combination of first-principles calculations and optical
pump-probe measurements. The article is organized as follows. In Section
\ref{sec:general} we briefly summarize the experimental observations that have
led to the conflicting viewpoints mentioned above and propose a working
definition for an \textit{exciton condensate} and \textit{chiral charge
density wave} that reconciles these phenomena. In Section \ref{sec:results} we
present our own experimental and theoretical results that demonstrate the
chiral CDW, which has been observed in optical pump-probe measurements, can be
explained by accounting for structural distortions that are screened by the
large electronic temperature that occurs in such pump-probe studies. In
particular, our working definition of an excitonic insulator (EI) that we
present in Section \ref{sec:name} and our results in Sec.~\ref{sec:results}
shows that (1) the chiral CDW in TiSe$_{2}$ is \textit{not} due to a
transition to an excitonic insulator phase and (2) the chiral optical
transitions observed in pump-probe studies are consistent with a transition of
the centrosymmetric (2$\times$2$\times$2) structure to a non-centrosymmetric
(2$\times$2$\times$1) structure, where these two structures are
near-degenerate in energy based on our first-principles calculations.

\section{General Considerations}

\label{sec:general}

\subsection{Summary of prior experimental studies}

Pump-probe optical measurements are often used to address the question of
whether the CDW transition is driven by an instability in the electronic or
ionic response, since in this experiment one can heat the electron subsystem
rapidly, and probe a combination of hot electrons and cold ions. Recently this
method was applied to another putative EI, Ta$_{2}$NiSe$_{5}$
\cite{baldini2020spontaneous}, where it was conclusively shown that the gap
opening is driven primarily by structural distortions. Numerous attempts to
use similar spectrosopic techniques on TiSe$_{2}$
\cite{lioi2016ultrafast,mohr2011nonthermal,ishioka2010chiral,
rohwer2011collapse,burian2020structurally} have also been reported.
Time-resolved x-ray diffraction (XRD) measurements or measurements of coherent
phonon oscillations
\cite{burian2020structurally,mohr2011nonthermal,rohwer2011collapse} performed
during these pump-probe studies have shown the structural distortion
associated with the CDW can be quenched as a function of increasing laser
fluence at lattice temperatures that are well below T$_{\mathrm{CDW}}$. Hence,
it was conjectured that since this transition is driven by increasing laser
fluence, this purportedly passes the test for TiSe$_{2}$ being an EI.
Furthermore, plasmon softening \cite{kogar2017signatures} has been measured at
T$_{\mathrm{CDW}}$ and was used as further evidence to support the EI nature
of the transition. However, we will show that these assumptions are tenuous.

A phenemenological description of the transient response upon photo-excitation
observed in these experiments can be understood as follows. The laser pump
excites electrons to a higher energy where the energy is equal to the photon
energy of the pump laser. Within a short timescale (femtoseconds) the
photoexcited electrons thermalize via electron-electron interactions, which in
turn raises the electronic temperature, $T_{e}$, of the sub-system while
the lattice temperature remains approximately fixed. Increasing the laser
fluence raises $T_{e}$. Thermalization with the lattice occurs over a
relatively longer time scale \cite{allen1987theory}.

The observation of chiral optical transitions at lattice temperatures around
or below the CDW transition temperature, $T_{\mathrm{CDW}}$, of 180 K
 is another piece of intrigue around TiSe$_{2},$
albeit not directly related with the putative EI physics. In this connection,
it was particularly exciting to see several reports of chiral optical
transitions \cite{xu2020spontaneous,lioi2016ultrafast,ishioka2010chiral},
\textit{i.e.,} optical transitions with finite circular polarization. 

Xu \textit{et al.} \cite{xu2020spontaneous} observed evidence of a chiral CDW
at and below 174 K through measurements of the circular photogalvanic effect (CPGE) current.
This finite CPGE signal occurs at a 
 a slightly lower temperature than $T_{\mathrm{CDW}}$, which they
attributed to a \textquotedblleft gyrotropic phase\textquotedblright\ with a
yet-to-be-determined non-centrosymmetric structure that is distinct from the
centrosymmetric (2$\times$2$\times$2) commensurate CDW phase.  
The CPGE is a second order nonlinear optical effect that is described by a third-rank tensor
that takes on a finite value when inversion symmetry is broken \cite{sipe2000second}.  To first order with
respect to an electric field, we will show this is equivalent to the off-diagonal components of the
dielectric tensor becoming finite for a hexagonal material such as TiSe$_2$. 
However, one key assumption within the study of Xu {\it et al.} \cite{xu2020spontaneous} is
 that the underlying atomic structure in the CDW phase already has
pre-existing chiral domains prior to photoexcitation that make the overall structure chiral.  In contrast, we will show
that for the case of pump-probe studies performed using high fluence, the CDW structure that is initially
centrosymmetric can take on a non-centrosymmetric structure up to a critical value of $T_e$.

Furthermore, at
a critical laser fluence and at a lattice temperature that is below
$T_{\mathrm{CDW}}$, the CDW and the finite chirality is quenched
\cite{lioi2016ultrafast,mohr2011nonthermal}, which has been
interpreted as a non-thermal melting of the CDW phase. 

\subsection{What's in a name?}

\label{sec:name} To interpret these experimental observations, it is
instructive to first consider a working definition for these two
phenomena;\textit{ i.e,} what is an \textit{excitonic insulator} and what is a
\textit{chiral charge density wave}?

When the term \textit{excitonic insulator} \cite{keldysh1965possible} was
first introduced, it was emphasized that the excitonic insulator is an
analogue of BCS superconductivity where the instability occurs in the
electron-hole (e-h), rather than the electron-electron (e-e) channel. Within this definition, one
would need to invoke higher-order interactions, such as ladder diagrams of the
Coulomb interaction to theoretically describe an excitonic insulator phase.
These ladder diagrams are not included in standard density functional theory
or in its Hartree-Fock like modifications. Hence, the advantage of this
definition is that it provides means to directly test whether a material
should be classified as an excitonic insulator based on the ability for
standard DFT to describe the physical observables associated with such a
phase.  Note that while any failure of DFT does not imply evidence of an
excitonic insulator state, the phenomena associated with an excitonic insulator
cannot be described by standard DFT.

For the case of TiSe$_{2}$ this includes a simultaneous description of the
(2$\times$2$\times$2) reconstruction of the lattice in the CDW phase, the
opening of a gap, the observation of plasmon softening at the CDW transition
\cite{kogar2017signatures} and non-thermal melting of the CDW in pump-probe
measurements at a lattice temperature well below the CDW transition
temperature \cite{rohwer2011collapse, mohr2011nonthermal,lioi2016ultrafast}.
Hellgren \textit{et al.} \cite{hellgren2017critical} used our definition of an
excitonic insulator (without explicitly stating this definition) and
demonstrated that hybrid functional DFT calculations, which do not incorporate
electron-hole ladder diagrams, are able to reproduce the observed (2$\times
$2$\times$2) commensurate CDW structure and to describe the insulating state
of TiSe$_{2}$.

It is worth noting that another seminal paper \cite{jerome1967excitonic},
which appeared two years after the work of Keldysh and Kopaev, implies that
any sort of electronic instability that does not require any lattice degrees
of freedom can be classified as an EI. This point of view was adapted in some
recent publications \cite{mazza2020nature,mohr2011nonthermal} to describe an
EI phase. We do not find it to be constructive, since this definition would
include any and all instabilities associated with a divergence in the plain
RPA dielectric function, such as a Peierls transition
\cite{peierls1955quantum}. Such instabilities predate the notion of an EI by
many decades and are not intuitively associated with an EI. Within this
interpretation it is assumed that the transition occurs entirely within the
electronic subsystem, while the ions follow the electronic CDW. While this
line of reasoning may seem compelling, one needs to consider the fact that the
ion-ion interaction is screened by electrons, and the response of the
electrons may (but does not have to) depend on the electronic temperature, and
thus can weaken or eliminate an ionic instability. If this mechanism is
operative, the CDW disappears simply because hot electrons in a narrow-gap
semiconductor, or in a semimetal, screen better than cold electrons, and
therefore suppresses the magnitude of the ion-ion interaction responsible for the instability.
Needless to say, this effect is not related to the notion of an excitonic insulator.

Regarding the \textit{chirality} of the CDW phase, a crystal structure is
chiral if it can be distinguished from its mirror image; that is, the latter
cannot be superimposed onto the original structure by any sequence of
rotations or translations \cite{kelvin1894molecular}. The crystal structure of
TiSe$_{2}$ is well established \cite{di1976electronic}, and indeed the CDW in
a single monolayer of TiSe$_{2}$ is chiral, that is, it does not possess a
center of inversion and it can formally support magnetooptical effects that
need broken time-reversal symmetry. However, the unit cell of the bulk
TiSe$_{2}$ in both the high temperature phase and the bulk CDW phase is
comprised of two monolayers of TiSe$_{2}$ stacked alternatively, so that sign
of the chirality (\textquotedblleft handedness\textquotedblright) alternates
and the crystal as a whole is achiral. It is then interesting to note that
chiral optical transitions have been associated with this centrosymmetric CDW
structure. However, these considerations do not preclude the presence of
symmetry-breaking structural distortions of the Ti and/or Se atoms that would
lead to a non-centrosymmetric structure.

So, for the purpose of this paper we will define a chiral CDW as a structure
that breaks inversion symmetry, and upon including spin-orbit coupling (SOC),
leads to a combination of broken inversion and time-reversal symmetry and in
turn, non-zero optical chirality. Formally, the observation of finite optical
chirality does not have to be necessarily related to breaking of spatial
symmetry breaking. For instance, one can imagine, theoretically, a situation
when a finite magnetization is generated by optical pumping (even though it is
rather unlikely in this material). Obviously --- and consistent with the
experiment, --- in TiSe$_{2}$ it is highly unlikely, particularly given that
the effect appears only upon heating up the electronic subsystem.

\section{Results}

\label{sec:results}

\subsection{Pump-probe measurements}

\label{sec:pump} We measure the transient change in the reflectivity, $R$,
that is induced by our pump pulse. Our experiments are conducted at 3 K with a
pump fluence of 0.17 mJ/cm$^{2}$. The circular dichroism (CD) in transmission
geometry is defined as $CD=\frac{A_{+}-A_{-}}{(A_{+}+A_{-})/2}$ where
$A_{+(-)}$ is the absorbance of the right (left) circularly polarized light.
However, within the reflection geometry the beam should have no CD signal at
normal incidence due to symmetry considerations. However, if the incident beam
is at a finite angle from the normal to the sample surface, then circular
dichroism in reflection geometry becomes finite.~\cite{silverman1986}

\begin{figure}[h]
\includegraphics[width=8.5cm]{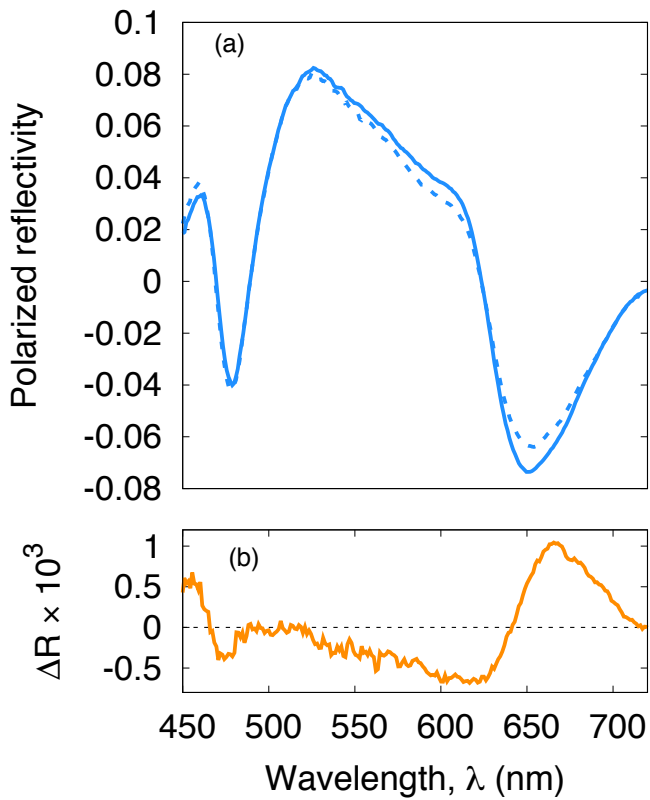}\caption{(a) Change in the right handed
(solid line) and left hand (dotted line) transient reflectivity obtained on
picosecond time scales as a function of wavelength. (b) Difference between the
right and left handed transient reflectivity, $\Delta R$, which is the degree
of circular polarization.
}%
\label{fig:expt}%
\end{figure}

Measurement of the CD using broadband transient optical reflectivity proceeds
by recording the transient change in reflectivity, $R$, induced by the
linearly polarized pump pulse, using a circularly polarized white light probe beam
(with right ($+$) and left ($-$) handed circular polarization) defined as:
$A_{+(-)}$ $(\lambda)$ = $-$ \textrm{lg} $\frac{R_{\mathrm{pump}}(\lambda
)}{R_{\mathrm{no-pump}}(\lambda)}$ The measured transient circular dichroism
(TRCD) signal in broadband time-resolved optical reflectivity experiment is
then $\Delta R = A_{+}(\lambda)-A_{-}(\lambda)$, which corresponds to the degree of
circular polarization. We identify two different relaxation time scales for
the TRCD: a picosecond time scale which is commensurate with electron-phonon
relaxation times and a longer nanosecond timescale that is typically
associated with phonon-phonon relaxation times.
Figure ~\ref{fig:expt} illustrates the wavelength dependence of the optical
density due to left and right circularly polarized light measured on the
picosecond time scale and the circular dichroism as a function of wavelength.

At 3 K (well below $T_{\mathrm{CDW}}$ $\sim$ 180 K), we find the degree of
circular polarization is finite in the spectral range between $\lambda$=550 nm
and $\lambda$=725 nm. We also find the degree of circular polarization to be
finite in this wavelength range for TRCD measured on the nanosecond timescale \cite{SM}. Further details on
these two relaxation time scales and the approach used to extract these decay
times from the raw transient reflectivity data are contained in
Ref.~\onlinecite{lioi2016ultrafast}. Our results, in agreement with Ref.
\cite{mohr2011nonthermal}, shows that above the critical fluence we use in
this study, the chirality of the optical transitions is zero and the chiral
CDW is quenched.

As we discuss in Sec.~\ref{sec:general}, the pump pulse used in our study will
lead to a change in the electronic temperature temperature $T_{e}$ of the
system while the lattice temperature remains fixed. Let us estimate what value
of $T_{e}$ this corresponds to. For a given photon fluence, $P$, material
volume, $V$, penetration depth of the excitation, $l$, electronic specific
heat for formula unit, $C_{e}$ and the magnitude of the reflectivity, $R$, the
critical electronic temperature is defined as $T_{e}=(1-R)PV/(lC_{e})$. The
electronic specific heat, $C_{e}$, is defined as: $C_{e}=\frac{\pi^{2}}%
{3}NT_{e}$, where $N$ is the average electronic density of states per formula
unit. The average electronic specific heat, $C$, when the TiSe$_{2}$ carriers
are heated from an electronic temperature of 0 K to $T_{0}$ K can defined as
$\frac{\pi^{2}}{3}N\int_{0}^{T_{0}}TdT=\frac{\pi^{2}T_{0}}{6}$ Hence,
 $T_{e}=6(1-R)PV/l\pi^{2}$. In our
experiment we use an excitation wavelength, $\lambda$, of 800 nm. The
absorption index at this wavelength is $\sim$ 3.2
\cite{buslaps1993spectroscopic,bayliss1985reflectivity}. Combining this
information, we determine the penetration depth, $l$ as $\lambda/(4\pi\kappa)$
and obtain a value of $l\sim$ 19 nm. The fluence, $P$, used for the results
report in Fig.~\ref{fig:expt} is 0.17 mJ/cm$^{2}$. Using the density
of states at the Fermi level of $\sim$ 1 state/eV per formula unit, from
our first-principles calculation, we find $C_{e}\sim2.36T_{e}$. The volume of
the TiSe$_{2}$ unit cell is 0.072 nm$^{3}$ and the reflectivity, $R$ at
$\lambda$=800 nm is $\sim$ 0.5 \cite{velebit2016scattering}.

Combining all of this information leads to a $T_{e}\sim$690 K. Note that
other studies have reported different values of fluence and a slightly
different $\lambda$ of 790 nm at which the CDW is suppressed. The values for
the critical fluence, $P$, ranges between 0.17 mJ/cm$^{2}$ to 0.5 mJ/cm$^{2}$
\cite{rohwer2011collapse, mohr2011nonthermal,lioi2016ultrafast}. This leads to
a range of values for $T_{e}$. the lowest value being $\sim$690 K and the
highest value is $\sim$1180 K.

The experimental data above shows that TiSe$_{2}$ exhibits a finite chirality
in the optical reflectivity with the peak in the chirality centered at $\sim$
625 nm. The lattice temperature for these measurements is 3 K while the
electronic temperature at the fluence used in Fig.~\ref{fig:expt} corresponds
to $T_{e}$ of 690 K. As we discuss in Sec.~\ref{sec:general} at a lattice
temperature of 3 K, TiSe$_{2}$ is expected to take on the centrosymmetric $P\bar{3}c1$
structure which should not lead to finite chirality. In the following we aim
to rationalize this puzzling observation by performing first-principles
calculations based on density functional theory (DFT).

\subsection{DFT calculations}

\label{sec:dft} Above the CDW transition temperature, bulk 1T-TiSe$_{2}$ is
stable in a hexagonal centrosymmetric structure (space group 164, $P\bar{3}%
m1$) where the Ti atoms are octahedrally coordinated by Se. We use the
experimental lattice constants of bulk TiSe$_{2}$ in the normal phase
($a$=3.527\AA ~and $c$=5.994\AA ) \cite{di1976electronic} to determine the
electronic structure. We find the high-$T$ phase to be a semi-metal, with a
hole-like pocket at $\Gamma$ and an electron-like pocket at \textrm{M} and
\textrm{L}, in agreement with previous calculations of the bulk structure of
TiSe$_{2}$ \cite{hellgren2017critical}. The hole pocket at $\Gamma$ is derived
primarily from Se $p_{z}$ states, while the electron pockets are derived from
Ti $d$-states.


The CDW transition temperature, $T_{\mathrm{CDW}}$, of bulk TiSe$_{2}$ is
$\sim$ 180 K \cite{di1976electronic,joe2014emergence}, below which TiSe$_{2}$
undergoes a ($2\times2\times2$) reconstruction to the commensurate CDW phase.
This is accompanied by a displacement of the Ti along the basal plane and a
minor rotation of the Se atoms around each Ti atom, \cite{di1976electronic}
and the corresponding space group of the structure changes from $P\bar{3}m1$
to $P\bar{3}c1$. Recent studies \cite{wegner2019local,wegner2020evidence} have
also suggested that displacements of the Ti and Se atoms that are
distinct from the $P\bar{3}c1$ structure. Within our DFT
calculations, we find each of these displacement patterns of the Ti and Se
atoms within the (2$\times$2$\times$2) reconstruction to be near degenerate in
energy with respect to each other. More importantly, all these structures are
centrosymmetric, similar to the $P\bar{3}m1$ unreconstructed structure or the
$P\bar{3}c1$ CDW structure.

\begin{figure}[h]
\includegraphics[width=7.5cm]{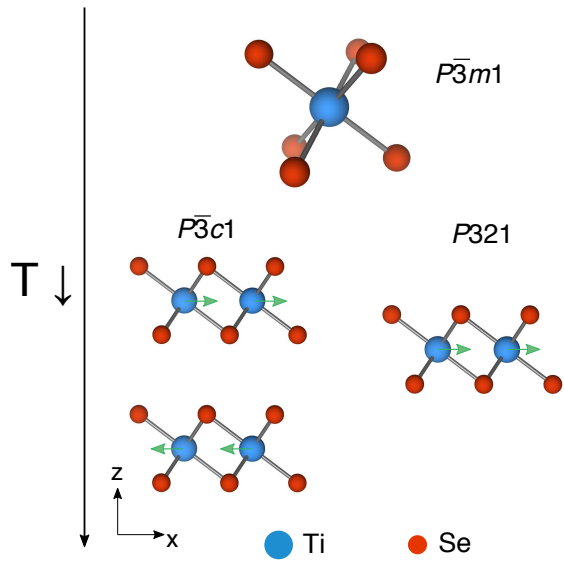}\caption{ Schematic depiction of how
the achiral centrosymmetric high temperature $P\bar{3}m1$ structure can
transform into either the achiral centrosymmetric $P\bar{3}c1$ CDW structure
or the chiral non-centrosymmetric $P321$ CDW structure as the temperature,
$T$, is lowered below the CDW transition temperature. The green arrows
indicate the direction that the Ti atoms are displaced in the CDW phase.}%
\label{fig:displacement2}%
\end{figure}

As we discuss in Sec.~\ref{sec:general}, a structural distortion that leads to
a non-centrosymmetric structure may be a plausible source of the finite
optical chirality that has been observed in the CDW phase. If we examine the
calculated phonon dispersion for TiSe$_{2}$ \cite{hellgren2017critical}, we
note that there are two soft modes, one at the \textrm{M} and a second at the
\textrm{L} high-symmetry points, which correspond to structural instabilities.
Indeed, the soft mode at the \textrm{L}-point corresponds to the (2$\times
$2$\times$2) CDW reconstruction. The soft mode at the \textrm{M}-point would
correspond to a (2$\times$2$\times$1) structural distortion. Such a structure,
which has a lower space group, $P321$, does not possess a center of inversion.
Figure \ref{fig:displacement2} illustrates these two possible structural
distortions that can occur starting from the high-temperature $P\bar{3}m1$
structure. 

In order to elucidate whether these structural distortions are impacted by
the elevated electronic temperatures that occur in pump-probe studies, we
performed DFT calculations, varying the magnitude of the electronic
temperature, $\sigma=kT_{e}$ from 0.005 eV to 0.1 eV (corresponding to an
effective $T_{e}$ of $\sim$58 K to $\sim$1160 K) and optimize the atomic
coordinates of TiSe$_{2}$ using both (2$\times$2$\times$1) and ($2\times
2\times2$) reconstructions. Within this approach we assume the photoexcited
electrons in our pump-probe study have thermalized (within femtoseconds as discussed
in Sec. \ref{sec:general}) to an electronic temperature that is determined in part
by the fluence while the lattice temperature remains fixed.

\begin{figure}[h]
\includegraphics[width=8.5cm]{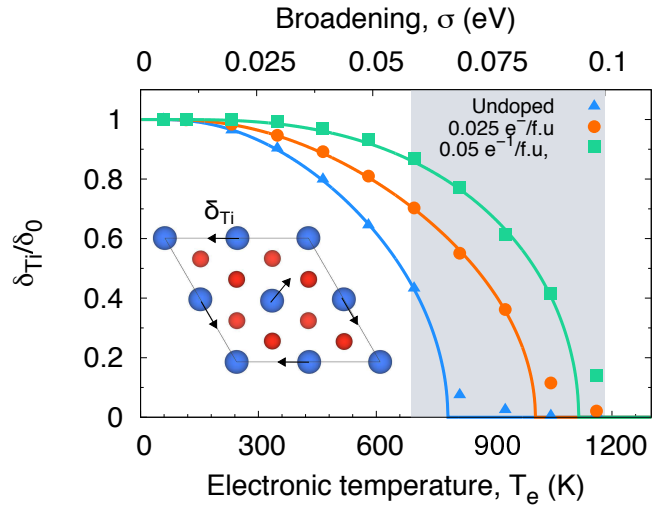}\caption{ Normalized displacement of the
Ti atoms in the $P321$ structure as a function of the magnitude of the
broadening, $\sigma$, and corresponding electronic temperature, $T_{e}$.
Results for an undoped (blue - $\bigtriangleup$), and doping with 0.025
$e^{-}$/f.u (orange - $\circ$) and 0.05 $e^{-}$/f.u (green - $\square$) are
illustrated. The solid lines are a fit to Eq.~\ref{eq:bcs}. The critical
temperature, $T_{c}$ where the structure transforms to $P\bar{3}m1$ from
$P321$ is the value of $T_{e}$ when the solid line intersects the horizontal
axis for each fit. The range of critical electronic temperatures where the
chirality of the CDW is found to be suppressed in pump-probe experiments is
illustrated with the grey shaded rectangle. A top view of the $P321$ structure
is illustrated in the inset with the arrows illustrating the direction the Ti
atoms are displaced.}%
\label{fig:displacement}%
\end{figure}

In practice, this is done by varying the magnitude
of the Fermi-Dirac energy broadening, $\sigma$, used in the self-consistent
cycle of our DFT calculations (cf. Sec.~\ref{sec:methods} and
Ref.~\onlinecite{SM}). For each value of $\sigma$, we perform a structural
optimization and determine the distance, $\delta_{\mathrm{Ti}}$, by which the
Ti atoms are displaced away from their their positions within the $P\bar{3}m1$
structure. In agreement with published
results \cite{hellgren2017critical}, we find the ($2\times2\times2$) to be the
ground state, and the magnitude of the displacement, $\delta_{\mathrm{Ti}}$ to
be somewhat underestimated (it was shown in Ref. \cite{hellgren2017critical}
that this may be corrected by adding a small fraction of Hartree-Fock exchange
using a hybrid functional). The normalized displacements with respect to the
displacement determined for the lowest energy broadening ($\sigma$ = 0.005 eV,
$T_{e}$=58K), $\delta_{0}$, are shown in Fig.~\ref{fig:displacement} for the
$P321$ structure.

We find that for low values of $\sigma$, the Ti atoms are displaced strongly
away from their corresponding high-symmmetry position and the structure
retains the low-symmetry $P321$ structure. However, at a critical value
of the electronic temperature,
$T_{e}=T_{c}$, the Ti atoms converge to their high-symmetry positions and the
structure is stable in the undistorted $P\bar{3}m1$ structure. We plot
$\delta_{\mathrm{Ti}}/\delta_{0}$, which we define as the order parameter
for this structural transition as a function $T_{e}$
and find that it exhibits a BCS-like 
temperature dependence. If we fit our first-principles calculations in
Fig.~\ref{fig:displacement} to the following BCS expression
\begin{equation}
\frac{\delta(T)}{\delta_{0}}=\mathrm{tanh}\left(  b\sqrt{\frac{1}{T}%
-1}\right)  \label{eq:bcs}%
\end{equation}
we find a critical temperature, $T_{c}$, to be 782 K, at which the non-centrosymmetric
$P321$ structure transforms to the high-temperature centrosymmetric $P\bar{3}m1$
structure (with a Hartree-Fock correction added as in Ref.
\cite{hellgren2017critical}, this temperature would likely be slightly higher).

We also consider the effect that intrinsic doping may have on this structural
phase boundary. Several studies have shown as-grown TiSe$_{2}$ exhibits
$n$-type conductivity that is likely due to unintentional impurities or native
defects \cite{watson2019origin,moya2019effect,campbell2019intrinsic}, which
act as a source of excess electrons. To this end, we simulate the effect of
$n$-type doping by changing the number of valence electrons and adding a
compensating jellium background charge, and optimize the atomic coordinates
and the volume starting from the $P321$ structure for different values of
$\sigma$. We investigate the effect of the following doping concentrations;
0.025 $e^{-}$/TiSe$_{2}$ f.u, 0.05 $e^{-}$/TiSe$_{2}$ f.u. 
The change in $\delta_{\mathrm{Ti}}/\delta_{0}$ as a
function of $\sigma$ with respect to doping is also 
illustrated in Fig.~\ref{fig:displacement}. 
We fit the results of $\frac{\delta
_{\mathrm{Ti}}}{\delta_{0}}$ for the two different doping levels to
Eq.~\ref{eq:bcs} and find $T_{c}$ increases to 1005 K for 0.025 $e^{-}%
$/TiSe$_{2}$ f.u and $T_{c}$ is 1115 K for a doping concentration of 0.05
$e^{-}$/TiSe$_{2}$ f.u.

We also conducted a similar analyses as in Fig.~\ref{fig:displacement}, taking
into account different approximations within DFT \cite{SM} and find
qualitatively similar behavior. The order parameter always exhibits a similar
BCS-like temperature dependence and the $T_{c}$ obtained by fitting
$\frac{\delta_{\mathrm{Ti}}}{\delta_{0}}$ versus $T_{e}$ to Eq.~\ref{eq:bcs}
for the different approximations we tested is within 6$\%$ of the $T_{c}$ for
undoped TiSe$_{2}$ reported in Fig.~\ref{fig:displacement}.

\begin{figure}[h]
\includegraphics[width=8.5cm]{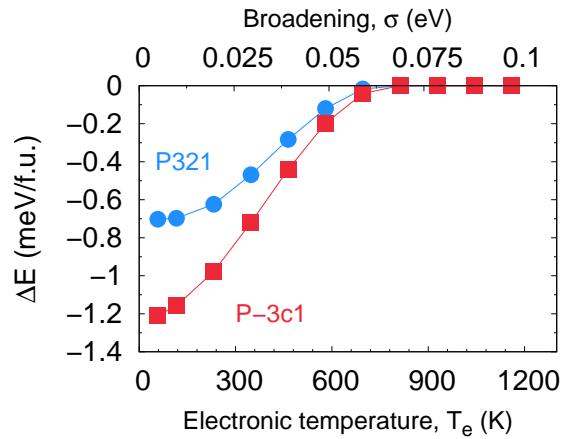}\caption{Change in total energy of the two
CDW structures, achiral $P\bar{3}c1$ ($\square$) and chiral $P321$ ($\circ$)
with respect to the total energy of the high-temperature $P\bar{3}m1$
structure as a function of the electronic temperature.}%
\label{fig:energy}%
\end{figure}

Applying the same procedure to the centrosymmetric $P\bar{3}c1$ CDW structure,
we find it also converges to the  $P\bar{3}m1$ structure at large values of $T_e$.
This is reflected in our calculations of the difference in the
total energy of the $P321$ and $P\bar{3}c1$ structures with respect to
$P\bar{3}m1$ structure as a function of $T_{e}$, as illustrated in
Fig.~\ref{fig:energy}. Note that while the $P\bar{3}c1$ structure is slightly
lower in energy than the $P321$ structure for all values of $\sigma$, in
agreement with the static equilibrium structure.  This
difference in energies is very small and may be reversed at some value of
$T_{e},$ once the effects of vibrational entropy are properly accounted for.
We speculate, based on our experimental findings, that this happens at some
electronic temperature, 
$T_{1}$, such that $0<T_{1}<T_{c}.$

Hence, from these calculations the following picture emerges. While the
centrosymmetric $P\bar{3}c1$ structure is the ground state CDW phase, the
non-centrosymmetric $P321$ structure is near-degenerate in energy. Hence, upon
photo-excitation, the rapid increase in $T_{e}$ stablizes the $P321$ structure
when $T_{e}$ becomes larger than $T_{1},$ and the magnitude of the displacement
of the Ti atoms with respect to the $P\bar{3}m1$ structure depends on $T_{e}$.
The $P321$ structure lacks a center of inversion and this, as we will show in
Section \ref{sec:optics}, leads to finite circular dichroism, \textit{i.e,} a
chiral CDW. When $T_{e}$ increases due to increasing fluence, the Ti atoms are
weakly displaced with respect to their positions within the $P\bar{3}m1$
structure. At $T>T_{c}$, the ion-electron interactions are sufficiently weak
so that the transition to the non-centrosymmetric $P321$, or the centrosymmetric $P\bar
{3}c1$ structure no longer occurs.  Instead, the achiral centrosymmetric $P\bar{3}m1$ structure is
stable and the finite circular polarization is quenched, i.e non-thermal
melting of the (chiral) CDW occurs.

Our estimate for the critical electronic temperature, $T_{c}$, from our
first-principles calculations, where the chiral $P321$ structure is quenched
ranges between 730 K for undoped TiSe$_{2}$ up to 1115 K for TiSe$_{2}$ doped
with 0.05 $e^{-}$/TiSe$_{2}$ f.u., which is well within the range of values
for the critical $T_{e}$ that we estimate by analyzing pump-probe experiments
at the end of Sec.~\ref{sec:pump}. This proves that the suppression of the
optical chirality as a function of increasing laser fluence observed in
pump-probe studies is not related to the presence of an excitonic insulator.
Instead it is due to structural distortions that are screened by an elevated
electronic temperature.

\subsection{Optical properties}

\label{sec:optics} 
\begin{figure}[h]
\includegraphics[width=8.5cm]{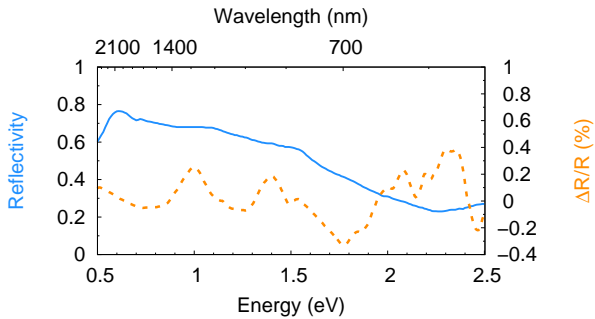}\caption{Reflectivity as a function
of photon energy (blue solid line - left vertical axis) and the degree of
circular polarization associated with these optical transitions (orange dotted
line - right vertical axis) calculated for the non-centrosymmetric $P321$
structure.}%
\label{fig:reflect}%
\end{figure}

To demonstrate that the centrosymmetric $P\bar{3}c1$ and the noncentrosymmetric
$P321$ structures do indeed lead to zero and non-zero chirality in their
optical transitions, respectively, we calculate the real and imaginary parts
of the dielectric function. We then determine the reflectivity, $R$, and the
degree of chirality, $\eta$ \cite{SM}. The calculated reflectivity is
illustrated in Fig.~\ref{fig:reflect}. 

First, we find the magnitude of the total reflectivity does not vary
significantly for either the $P\bar{3}c1$ or the $P321$ CDW structures or the
high-temperature $P\bar{3}m1$ structure. For photon energies between 0.5 eV
and 2.5 eV the magnitude of the reflectivity ranges between $\sim$0.4 and
$\sim$0.6. The reflectivity for the $P321$ structure is illustrated in
Fig.~\ref{fig:reflect}. We note that magnitude of the reflectivity across this
range of energies is also consistent with what has been measured
experimentally \cite{velebit2016scattering}.

To determine the degree of circular dichroism of the reflectivity we then
compare the difference between the off-diagonal components of the imaginary
part of the dielectric function \cite{SM}. Since the high temperature
$P\bar{3}m1$ structure and the $P\bar{3}c1$ CDW structure both possess a
center of inversion and they also preserve time-reversal symmetry, left and
right circularly polarized optical transitions are equal in magnitude and, as
a result, one would expect the degree of circular dichroism to be zero.
Indeed, we find this to be the case from our first-principles calculations for
the two centrosymmetric structures, i.e they are achiral. 

However, for the
non-centrosymmetric $P321$ structure, we find the off-diagonal components of
the dielectric function are finite and this leads to finite circular
polarization as we show in Fig.~\ref{fig:reflect}. The degree of circular
polarization is finite within an energy range of 1.5 eV to 2 eV, which is
approximately the energy range where finite circularly polarized transitions
occurs in our pump-probe measurements of TiSe$_{2}$
(Fig.~\ref{fig:expt}(b)). This is also consistent with the range of
wavelengths where finite circular polarization has been observed in previous
pump-probe studies of TiSe$_{2}$ \cite{rohwer2011collapse, mohr2011nonthermal}.
We also conjecture that continuous wave excitation under sufficiently high
power, long duration and at lattice temperatures below $T_{\rm CDW}$, 
as used in the study by Xu {\it et al.} \cite{xu2020spontaneous}, may also
conspire to lead to the finite CPGE current.  
Based on our results and discussion presented above, this does not require the
pre-existence of chiral domains.  Instead, photoexcitation above a critical power
may lead to a finite fraction of the atomic structure to be trapped in the 
non-centrosymmetric $P321$ structure, which would lead to finite chirality.

Hence, from these calculations we can conclude the following. The observation
of chiral optical transitions during pump-probe measurements on TiSe$_{2}$
cannot be explained by the conventional $P\bar{3}c1$ structure that is
associated with the (2$\times$2$\times$2) commensurate CDW phase of TiSe$_{2}%
$. Instead, a symmetry breaking mechanism that leads to a finite difference
between right and left circularly polarized optical transitions has to be
operative. Our calculations suggest the non-centrosymmetric $P321$ structure
that is near-degenerate in energy with the $P\bar{3}c1$ is stabilized upon
photo-excitation. The $P321$ structure leads to finite circularly polarized
transitions that are consistent with the signatures of the chiral CDW that
have been identified in pump-probe studies of TiSe$_{2}$.

\section{Conclusions}

Comparing our first-principles calculations with experiment, we 
draw two main conclusions. First, the experimentally observed non-thermal
melting of the CDW in TiSe$_{2}$, that is, melting of the CDW upon heating the
electron subsystem, can be quantitatively explained by the effect of the
electronic temperature on the electron screening of the ion-ion interactions.
This is a one-electron effect not related in any manner with the physics of
excitonic insulators. Therefore, non-thermal melting of the CDW alone does not provide
evidence for the existence of an excitonic insulator state in TiSe$_{2}.$

Second, we observe, in agreement with previous measurements, that the chiral
optical response of TiSe$_2$ exists for a finite range of laser fluences, in other words,
within a finite range of electronic temperatures. To this effect, we calculated
the energy difference between the ground state centrosymmetric CDW structure ($P\bar{3}c1)$ and the
non-centrosymmetric $P321$ CDW structure, and found it to be extremely small ($\sim10$
K).  Furthermore, this energy difference rapidly decreases as the electronic temperature increases.
We conjecture that there
is a small additional energy term, possibly related to vibrational entropy, that is
outside the scope of our static lattice DFT calculations, which
is either independent of the electronic temperature, or even grows with it.
Such a contribution would impact the energy difference between the $P321$ and $P\bar{3}c1$
structures and lead to the noncentrosymmetric $P321$ structure to be the ground state
at some intermediate electronic temperature, $T_1$.

\section{Methods}

\label{sec:methods}

\subsection{Experiments}

Single crystals of TiSe$_{2}$ were prepared using the chemical vapor transport
technique with iodine as a transport agent in a manner similar to that used by
Oglesby \textit{et al.} \cite{oglesby1994growth}. We used Titanium powder
(99.99\%), Selenium powder (99.999\%), and Iodine powder (99.99\%) purchased
from Alfa Aesar. The powders were premixed and inserted into quartz tubes
which were evacuated to 10$^{-4}$mTorr. We allowed crystal growth to proceed
for 3 weeks before quenching the reaction and retrieving the single crystals.

Broadband transient optical reflectivity was measured using the split output
of an 800 nm 35 fs Ti:sapphire laser system equipped with an optical
parametric amplifier. The pump pulses, having a wavelength of 800 nm (1.55
eV), were incident on the sample at 1 kHz using a synchronized chopper that
blocks every other pulse of the parent laser operating at a 2 kHz repetition
rate. A delayed white light supercontinuum probe was generated by focusing the
800 nm laser into a CaF$_{2}$ plate. The pulse beam had a 600~$\mu$m diameter
spot size and the probe beam spot size had a 100~$\mu$m diameter. An
adjustable time delay is established between the pump and probe pulse by means
of a mechanical track that adjusts the path of the probe pulse. The pump beam
was incident on the sample at an angle of incidence close to 20$^{\circ}$ and
the probe beam was oriented with an angle of incidence at 30$^{\circ}$. The
pump pulse is linearly polarized while the probe pulse is circularly polarized
(CP). In order to control the handedness of the CP probe beam a Pockels cell
was inserted into the probe beam path. The Pockels cell using a KD*P crystal
was driven using the low jitter Q-switch driver QBU-BT-6024-LJ from Vigitek,
Inc. (http://vigitek.biz). All of the pump-probe data was obtained at 3 K
(well below T$_{\mathrm{CDW}}$ = 200 K) with a pump fluence of 0.17
mJ/cm$^{2}$.

\subsection{First-principles calculations}

Our calculations are based on density functional theory within the
projector-augmented wave method \cite{Blochl_PAW} as implemented in the Vienna
\textit{Ab-initio} Simulation Package (VASP) \cite{VASP_ref,VASP_ref2}. All of
the results in the main text use the generalized gradient approximation (GGA)
defined by the Perdew-Burke-Ernzerhof functional \cite{perdew1996generalized}.
We use the Ti PAWs where the $3d$, $4s$, $4p$ and the Se PAWs where the $4s$,
$4p$ electrons are treated as valence and a plane-wave energy cutoff of 400
eV. All of the structural relaxations of the bulk unit cell used a (24$\times
$24$\times$12) $k$-point grid. Calculations of the (2$\times$2$\times$2) and
the (2$\times$2$\times$1) CDW structures used $k$-point grids that were scaled
with respect to the k-point grid used for the unit cell. An energy convergence
criteria of 10$^{-8}$eV and a force convergence criteria of 2
meV/$\mathring{A}$~was used for all of the calculations. To examine the
structural phase transition as a function of electronic temperature we used
the Fermi-Dirac smearing scheme. The Grimme-D3 correction scheme was used to
account for van-der-Waals interactions \cite{grimme2010consistent}. The space
groups of the different structures were determined using {\rm spglib}
\cite{togo2018texttt}. To determine the circular polarization of the optical
transitions, we calculate the imaginary and real part of the dielectric
function with spin-orbit interaction included. 

\section{Data availability}
The data that support the findings of this study are available from the corresponding author upon reasonable request.

\acknowledgements
We thank Nuh Gedik, Qiong Ma, and Su-Yang Xu for insightful discussions.
D.W. was supported by the Laboratory-University Collaboration
Initiative of the DoD Basic Research Office.
G.K acknowledges support from the NSF under
Grant No. ECCS-1711015. Use of the Center for Nanoscale Materials, an Office
of Science user facility, was supported by the U.S. Department of Energy,
Office of Science, Office of Basic Energy Sciences, under Contract No. DE-AC02-06CH11357.
 I.I.M. was
supported by ONR through grant N00014-20-1-2345.

\end{document}